\documentclass[fleqn, usenatbib]{mnras}

\usepackage{graphicx}	
\usepackage{amsmath}	
\usepackage{amssymb}	
\usepackage{multicol}   
\usepackage[T1]{fontenc}
\usepackage{newtxtext, newtxmath}

\title[Merging Compact Dwarf Galaxies]{Forming Blue Compact Dwarf Galaxy (BCD) through Mergers}

\author[D. N. Chhatkuli et al.]{
Daya Nidhi Chhatkuli,$^{1}$
Sanjaya Paudel,$^{2}$\thanks{E-mail:sanjpaudel@gmail.com}
Rajesh Kumar Bachchan,$^{3}$
Binil Aryal,$^{1}$\newauthor
Jaewon Yoo$^{4,5}$
\\
$^{1}$Central Department of Physics, Tribhuvan University, Kirtipur, Kathmandu, Nepal\\
$^{2}$Department of Astronomy, Yonsei University, Seoul, 03722, Republic of Korea\\
$^{3}$Patan Multiple Campus, Tribhuvan University, Lalitpur, Nepal\\
$^{4}$Quantum Universe Center, Korea Institute for Advanced Study (KIAS), 85 Hoegiro, Dongdaemum-gu, Seoul 02455, Korea\\
$^{5}$Korea Astronomy and Space Science Institute (KASI), 776 Daedeokdae-ro, Yuseong-gu, Daejeon 34055, Republic of Korea
}

\date{Accepted \today. Received \today.; in original form \today.}

\pubyear{2022}

\begin{document}
\label{firstpage}
\pagerange{\pageref{firstpage}--\pageref{lastpage}}
\maketitle

\begin{abstract}
It has long been speculated that Blue Compact Dwarf galaxies (BCDs) are formed through the interaction between low-mass gas-rich galaxies, but a few candidates of such systems have been studied in detail. We study a sample of compact star-forming dwarf galaxies that are selected from a merging dwarf galaxy catalog. We present a detailed study of their spectroscopic and structural properties. We find that these BCDs looking galaxies host extended stellar shells and thus are confirmed to be a dwarf-dwarf merger. Their stellar masses range between $8\times10^7$~M$_{\sun}$ and $2\times10^9$~M$_{\sun}$. Although the extended tail and shell are prominent in the deep optical images, the overall major axis light profile is well modeled with a two-component S\'ersic function of inner compact and extended outer radii. We calculate the inner and outer component stellar-mass ratio using the two-component modeling. We find an average ratio of 4:1 (with a range of 10:1 to 2:1) for our sample, indicating that the central component dominates the stellar mass with an ongoing burst of star-formation. From the measurement of H$_\alpha$ equivalent width, we derived the star-formation ages of these galaxies. The derived star-formation ages of these galaxies turn out to be in the order of a few 10 Myr, suggesting the recent ignition of star-formation due to events of satellite interaction.
\end{abstract}



\section{Introduction}\label{sec:intro}

\begin{figure*}
\includegraphics[width=17cm]{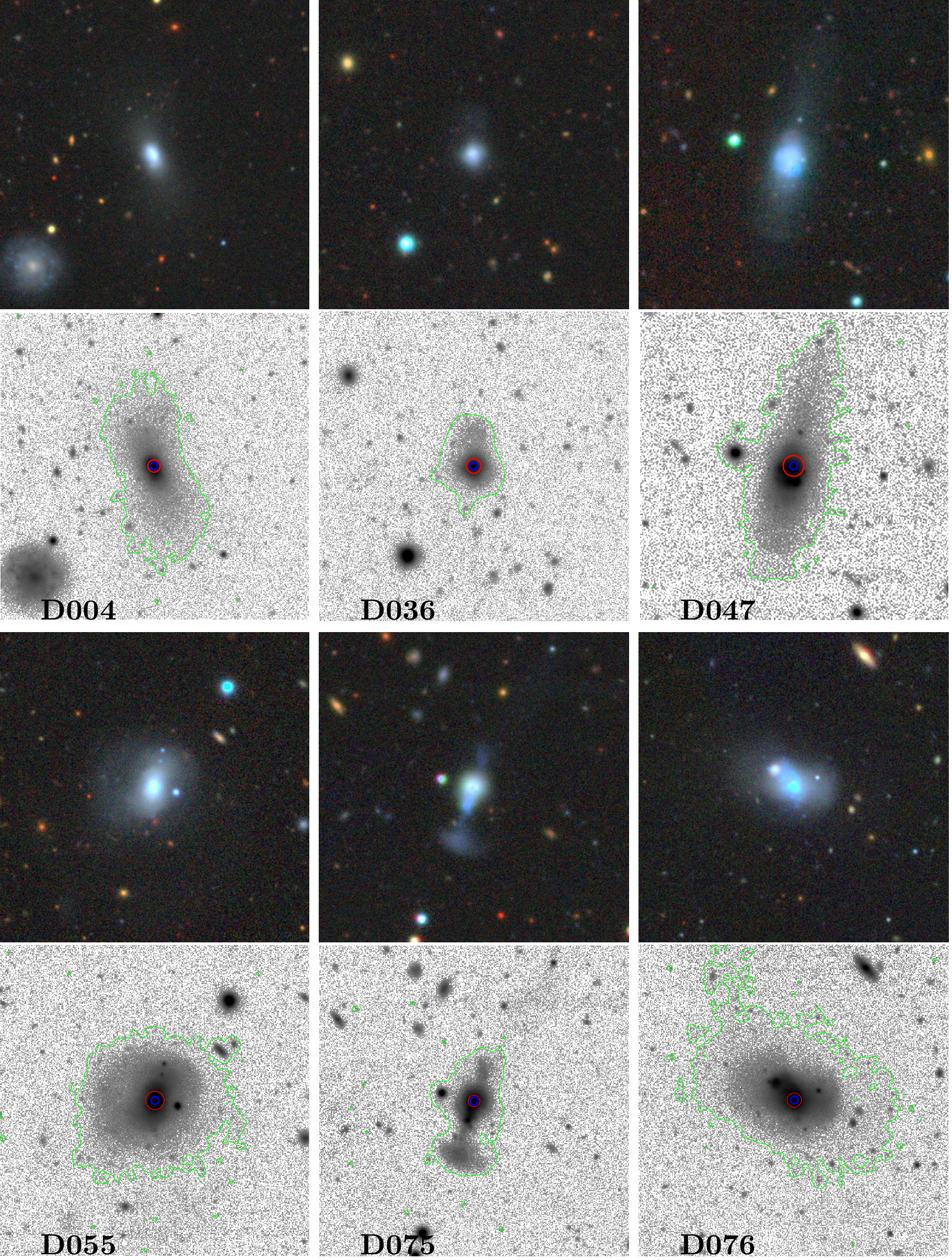}
\caption{Legacy survey color images of our sample galaxies. The second and fourth rows represent gray-scale $g$-band images with a field of view of $2\arcmin\times2\arcmin$. In the second and fourth rows, we show tri-color images of the respective galaxies, produced from a combination of $g-r-z$ band images.  We also show the size of the measured half-light radius with a red circle and a blue circle representing the SDSS fiber position. In each gray scale image, we draw a green contour representing a surface brightness level of 27 mag-arcsec$^{-2}$ at the $g-$band.  }
\label{main}
\end{figure*}

A significant effort has been made to study the interaction of massive galaxies, while mergers between dwarf galaxies ($M_{\star}$ $<$ 10$^{9}$) received little attention until very recently. Because dwarf galaxies have shallow potential-well, their evolution is expected to be more  driven by the large-scale environment than by merging events  \citep{Boselli06,Kormendy09,Lisker09}. Indeed, both observations and numerical simulations have extensively concluded that massive elliptical galaxies were likely to be originated first major mergers, building a dense inner core at high redshift, and grow through minor mergers, contributing to the formation of extended outer stellar halos \citep{Khochfar06,Oser10,Duc11}.

 In contrast to their massive counterparts, star-forming dwarf galaxies, and hence their merger, are often gas-dominated. 

Recent observations have revealed that sub-structures like a stellar shell, kinematically decoupled core, or even tidal stream around dwarf galaxies are common, which can be considered as the signatures of past merger events \citep{Geha05,Rich12,Penny12,Toloba14}. \citet{stierwalt15} carried out the first systematic study of gas-rich dwarf-dwarf interacting pairs and found that the star formation rate (SFR) of relatively close pair is enhanced by a factor of $\sim 2$ on average. \citet{pearson16} further showed that gas-rich dwarf pairs have more extended atomic gas distribution than unpaired analogs. In addition, case studies of several star-forming dwarf pairs have been carried out recently \citep{annibali16,privon17,Paudel17b,Paudel18b}.

Structural parameters such as size, concentration, and ellipticity are important tools to study galaxy formation and evolution. Galaxies follow a scaling relation between size and magnitude that is continuous over the whole range of luminosities \citep{Graham03,Ferrarese06,Janz08}. The universality of this relation has been a subject of considerable debate because of the presence of outliers and non-linearity of the relation \citep{Kormendy09,Chen10}.  In particular, early formed galaxies are compact compared to the later formed galaxies, as we see that the high red-shift galaxies are significantly compact compared to the local ($z < 0.1$) galaxies.  The size, evolution, and formation of compact galaxies in the early universe are some of the mysteries of current cosmology. Compact dwarf galaxies are prominent outliers of galaxies in the conventional size-magnitude relation \citep{Paudel14}. Their formation and evolution have been poorly understood. Compact early-type galaxies, like M32, are mainly found in the vicinity of large galaxies, e.g. M31, and such proximity to the nearby giant has led to an argument that the galaxies like M32 might have formed through the tidal stripping \citep{Faber83}. On contrary, recent discoveries are evidence that all compact dwarf galaxies are not located nearby of giant galaxies. \cite{Huxor13} found that a compact dwarf galaxy is located in an isolated environment. Such discovery of isolated compact dwarf galaxies has put forward another hypothesis that compact dwarf galaxies might have formed through the merging of even smaller dwarfs. Indeed, \cite{Paudel14b} discovered a nearly isolated compact dwarf galaxy, CGCG 036-042, with prominent evidence of merging features. 

\begin{table*}
\caption{Basic physical parameters of six BCDs.}
\centering
\begin{tabular}{ccccccccccc}
\hline
Galaxy & RA        &  DEC     &    $z$    & $D$ & $g-r$  & $M_{r}$  & $M_{*}$ & SFR  & $M_{\text{gas}}$ & Feature\\
\hline
 & ($\degr$) & ($\degr$) & & (Mpc) & (mag) & (log(M$_{\sun}$)) & (log(M$_{\sun}$)) & (log(M$_{\sun}$/yr)) & (log(M$_{\sun}$)) & \\
\hline
D004 & 028.9989 & -0.1855 & 0.0121 &  51.57 & 0.46 &  -16.72   & 9.09  &  -0.91   & 8.43 &  Sh \\
D036 & 144.5608 & 19.7111 & 0.0144 &  61.48 & 0.20 &  -16.46   & 8.56  &  ---     & 8.87 &  Sh \\
D047 & 150.3099 & 37.0709 & 0.0048 &  20.34 & 0.14 &  -15.94   & 8.26  &  -1.29   & ---- &  Sh \\
D055 & 157.4553 & 16.1809 & 0.0108 &  45.98 & 0.23 &  -17.89   & 9.18  &  -0.68   & 8.19 &  SL \\
D075 & 177.0757 & -1.6399 & 0.0130 &  55.44 & 0.22 &  -17.54   & 9.03  &  -0.51   & 9.94 &  Sh \\
D076 & 177.5113 & 15.0231 & 0.0024 &  11.90 & 0.09 &  -15.29   & 7.92  &  ---     & 8.40 &  Sh \\
\hline
\label{mtab}
\end{tabular}
\\
{The first column is the name according to the \cite{Paudel18a} catalog. Columns 2-10 represent RA, DEC, red-shift, Hubble flow distance of the galaxies, $g-r$ color, $r$-band magnitude, stellar mass, star-formation rate, HI gas mass, respectively. In last column we list the tidal feature of the galaxies as noted in Paudel et al.}
\label{sample}
\end{table*}

Blue Compact Dwarf galaxies (BCDs)  are low-luminosity galaxies ($M_{B} > -18$~mag) with a compact optical appearance and are blue in color \citep{Papaderos96}. In particular, super-compact BCDs are considered young objects which have accumulated most of their stellar mass at late cosmic epochs \citep{Drinkwater91}. Various lines of thought have been presented for their origin and future evolution. The mechanism that triggers the burst of star formation in these compact dwarf galaxies remains a mystery. Mergers, fly-by encounters, or gas turbulence have been proposed to explain the recent burst star-formation activity in BCDs \citep{Noeske01,Pustilnik01,Bekki08}.

Here, we present a study of a unique sample of  BCDs that recently underwent merging activity and hosts clear stellar shells and streams as a signature of a recent merger.

\section{Sample and Data Analysis}
To study the formation and evolution of BCD-type galaxies, we preferentially selected six compact merging dwarf candidates with a clear tidal feature. Our sample galaxies are taken from the catalog of merging dwarf galaxies prepared by \citet{Paudel18b}. The basic properties of our sample galaxies are listed in Table~\ref{mtab} that we have fetched from the parent catalog. These dwarf galaxies have a stellar-mass range of $8times10^{7}$~M$_{\sun}$ to $2\times10^{9}$~M$_{\sun}$, which are comparable to the Large Magellanic and Small Magellanic Clouds of the local group. By chance, we found a substantial amount of multi-wavelength data of these galaxies in public archives, which allowed us to thoroughly analyze their morphology, chemical properties, and stellar populations.
Figure~\ref{main} reveals the optical view of our galaxies, where we can see an elongated low surface brightness tidal tail along the major axis. The color image is a cutout of the Legacy sky-survey imaging database \citep{Dey19}, which is prepared by combining $g-$, $r-$, and $z-$band images. The field-of-view of the image is $2' \times 2'$. We can see that the inner part is quite round and bluish, and the outer low surface-brightness extended tidal tail is relatively reddish. We also show the size of the measured half-light radius with a red circle and a blue circle representing the SDSS fiber position. Given their compact nature, D036 and D075 are visually similar to tadpole galaxies identified by \citet{Straughn15} in Hubble Deep Field. D036 has a short faint tail-like structure that shows one-side elongation. On the contrary, D076 seems to have a tail on both sides, with one side's low-surface brightness tail quite prominent compared to the other. 

All, except D047, are located in a nearly isolated environment. The D036 is located at the edge of small group CGCG\,092-005 at a 445~kpc sky-projected distance from the group center.  

\begin{figure*}
\includegraphics[width=18cm]{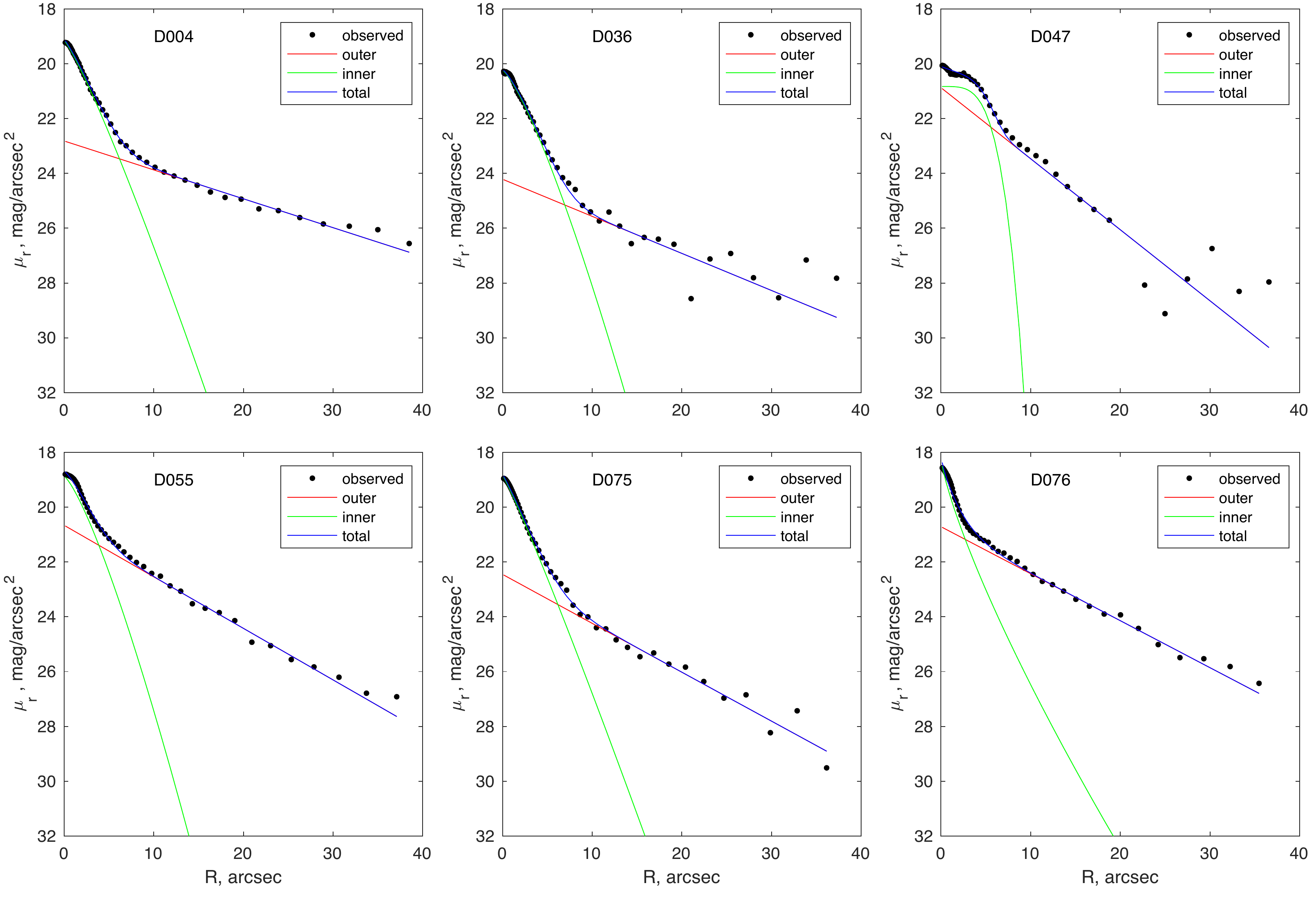}
\caption{Multi-component S\'ersic modelling of observed one-dimensional light profile. The black dots represent the observed data points. We show the inner and outer best fit S\'ersic function in green and red lines, respectively. The blue line represents a combination of both the inner and outer best-fit model.}
\label{serfit}
\end{figure*}

\subsection{Imaging and Photometry}
We used the SDSS image to perform image analysis \citep{Ahn12}. For this purpose, the SDSS $r-$band images have been used extensively because they have a higher signal-to-noise ratio than the other bands and we mainly used this band image to perform surface photometry. The surface photometry of very low surface brightness objects is notoriously difficult. There are several factors that are crucial in making an accurate estimation of photometric parameters.  Sky background is among the most important ones. Although the sky background subtraction in the latest version of the SDSS pipeline processed images is relatively good, we additionally process the SDSS images. We retrieved stacked images ($3\arcmin \times 3\arcmin$ fits-image cutout) from the SDSS data archival service (DAS). We subtracted the sky background preparing a background map for each object. The background maps were constructed after masking out all identified sources in the image, which were defined by source-extractor segmentation maps, and the segmentation images were filled by the median values calculated from all the surrounding pixels. This method allows us to eliminate any contribution of light from stars and background galaxies. Finally, the background map was subtracted from the original fits file to remove the sky background contribution to the observed flux.

To perform surface photometry, we exploit the Image Reduction and Analysis Facility (IRAF)  {\it ellipse} task, which outputs the major-axis profile of average intensity and ellipse parameters, such as position angle and ellipticity. The {\it ellipse} uses the methodology presented in \citet{Jedrzejewski87}, where for each semi-major axis length, an azimuthally averaged intensity $I(\phi)$  is calculated within the concentric ellipse area. The ellipse is defined with an initial guess for the isophote's center ($X, Y$), ellipticity ($\epsilon$), and semi-major axis position angle ($\theta$). The best fit ellipse is obtained by minimizing the sum of the squares deviation between the data while expanding the $I(\phi)$ into a Fourier series as:
\begin{equation}
I(\phi) = I_0 + \sum_k[A_k{\rm sin}(k\phi) + B_k{\rm cos}(k\phi)].
\end{equation}
and for a good approximation of the ellipse, the expansion should be truncated to the first two terms.

All foreground and background unrelated objects were masked. The masking was performed manually, where we visually identified the unrelated objects. The center of galaxies was derived by using {\it imcentre} task in the IRAF, which calculates a centroid of the provided image section. While fitting the ellipse, the position angle and the ellipticity were allowed to vary freely, and the semi-major axis was increased logarithmically. The ellipse center was not allowed to move by more than 3 pixels between the successive isophotes. 

The derived $r-$band major axis light profiles of our sample galaxies are shown in Figure~\ref{serfit}. The light profile generally shows a break which shows an abrupt change in gradient. We see that the inner component's light profile rises steeply compared to the outer component. However, in the case of D047, we see a somewhat irregular profile in the inner region. We then approximate the observed galaxy light profile into a multi-component S\'ersic function, namely $I_{\text{overall}} =  I_{\text{In}} + I_{\text{Out}}$, where the S\'ersic \citep{Sersic68} function is defined as
\begin{equation}
I(R)=I_{\rm e}\exp\left\{ -b_n\left[\left( \frac{R}{R_{\rm e}}\right) ^{1/n} -1\right]\right\}, \label{EqSer}
\end{equation}
where $I_{\rm e}$ is the intensity of the light--profile at the effective radius $R_{\rm e}$, and $n$ defines the `shape' of the profile.  The term $b_n$ is simply a function of $n$ and chosen to ensure that the radius $R_{\rm e}$ encloses half of the profile's total luminosity \cite{Graham05}. 

In Figure~\ref{serfit}, we show the modeled inner and outer components in green and red solid lines, respectively.  The overall model two-component function is shown in the blue line. To obtain the best fit model, we used $\chi^{2}$-minimization scheme. In doing so, we found that the outer components generally have a near exponential (n=1) S\'ersic index, and we decided to fix n = 1 for the outer component in the final fitting.

The two-component modeling is used to decompose merger remanent in its progenitor, assuming that each component roughly represents the individual progenitor. Finally, we performed integrated photometry in each component model to calculate each component's total flux and magnitude. We estimated the stellar masses by multiplying the SDSS-$r$-band fluxes with a mass-to-light ratio obtained from \citet{Zhang17} for the observed $g-r$ color. These values are most likely upper limits.

The stellar mass ratio between the two components of our sample BCDs varies between 0.1 to 0.5, where we calculate stellar mass from their $r$-band flux and $g-r$ color. However, we warn that there is a reasonable chance that both components' stellar populations are well mixed and therefore, the calculation may have a large uncertainty.

Because of the complexity in the observed light profile of the merging galaxies, no overall structural parameters can be derived using a simple S\'ersic function modeling. To measure the overall photometric properties, we used a non-parametric approach, i.e., the Petrosian method \citep{Janz08}. 

Using the Petrosian method, we estimated the value of total luminosities and half-light radii. The global half-light radii of our sample galaxies range from $0.23$~kpc to $0.56$~kpc which, indeed, proves that our selected sample dwarf galaxies are compact dwarfs in general. 

\begin{table}
\caption{Derived properties of the sample galaxies.}
\begin{tabular}{cccccccc}
\hline
ID &  $ c $     &   $Z$    & SFR & $R_{a}$ & $R_{i}$ & $n_{i}$ & $R_{o}$ \\
   &  & (dex) & log(M$_{\sun}$/yr) & ($\arcsec$) & ($\arcsec$) &  & ($\arcsec$)\\
\hline
D004  & 4.62 & 8.33 & 0.022  & 2.841 & 2.490 &  0.8 & 17.1  \\
D036  & 3.34 & 8.33 & 0.022  & 2.101 & 2.751 &  0.8 & 19.4  \\
D047  & 3.27 & 8.14 & 0.002  & 4.234 & 1.212 &  0.3 & 06.1  \\
D055  & 3.56 & 8.23 & 0.148  & 3.162 & 2.481 &  0.7 & 10.0  \\
D075  & 2.98 & 8.13 & 0.118  & 2.659 & 2.358 &  0.9 & 10.4  \\
D076  & 3.26 & 7.99 & 0.061  & 1.998 & 2.067 &  1.2 & 10.5  \\
\hline
\end{tabular}
\\
The first column is ID of the galaxies. The Balmer decrement, emission line metallicity and SFR derived from spectroscopic data are listed in columns 2, 3 and 4, respectively. In the columns 5, 6, 7 and 8, we list the overall galaxy size ($R_{a}$), inner size ($R_i$), S\'ersic index ($n_i$) and the outer size ($R_o$) respectively.
\label{partab}
\end{table}

\subsection{Spectroscopy}
All of our sample galaxies are targeted by the SDSS fiber spectroscopy observation. We retrieved the optical spectra of these galaxies from the SDSS data archive. These optical spectra have a reasonable signal-to-noise ratio. They are observed with the fibers spectrograph of  3$\arcsec$ diameter. The sample galaxies have a broad range of distances; the central 3$\arcsec$ represents a 50 pc to 100 pc core region depending on their distance from us. 

The SDSS spectra show strong emission lines, particularly Balmer lines, and resemble a typical spectrum of a star-forming H II region. The emission line fluxes were measured after subtracting the stellar absorptions, which are particularly strong in the case of H$_\beta$. For this, we used a publicly available code, the GANDLF \citep{Sarzi06} of a version particularly designed for the SDSS spectra, which uses a combination of stellar templet from \citet{Tremonti04}. The emission line fluxes are measured by fitting a gaussian profile which allows us to measure total flux, FWHM and equivalent width simultaneously. For this purpose, we use IRAF {\it splot} task. 

We derived Oxygen abundances ($12+\log$[O/H]) using these emission line fluxes. The extinction coefficient, $E(B-V)$, were derived from H$_\alpha$/H$_\beta$ emission-line ratios using a calibration provided by  \citet{Cardelli89}. The derived value of the extinction coefficient varies between 0.12~mag to 0.21~mag. D004 shows the largest extinction and D075 shows the least extinction (Table~\ref{partab}). We finally derived the SFR from the extinction corrected emission line H$_\alpha$ flux using the calibration of \citet{Kennicutt98}.

\section{Results}
\subsection{Imaging}
We have performed the surface photometry and the emission line analysis on a sample of merger remanent dwarf galaxies that are selected from a merging dwarf galaxy catalog. We used a publicly available optical imaging and spectroscopic data from the SDSS data server. The derived structural and spectroscopic parameters are listed in Table~\ref{partab}. As expected, the inner component is in general five times smaller than the outer component and has a smaller S\'ersic index. The measured photometric parameters show that the outer component is significantly redder than the inner component. The difference in $g-r$ color index of two-components is $\approx0.2$ on average. 

\begin{figure}
\includegraphics[width=7.5cm]{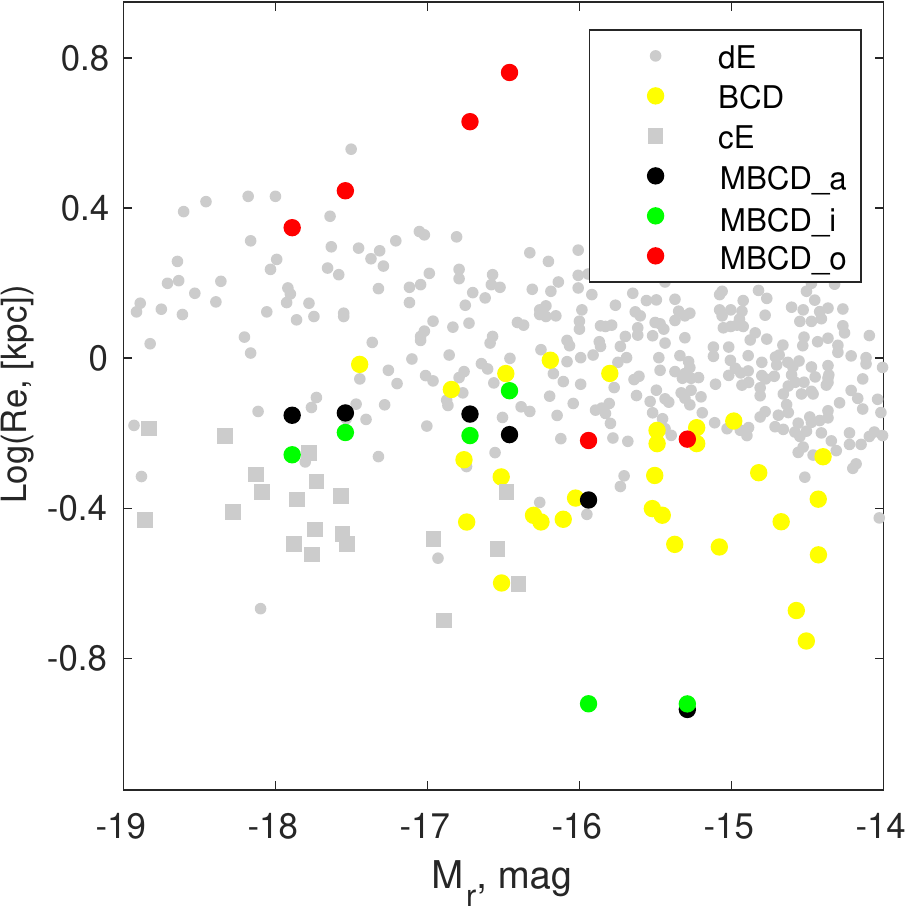}
\caption{Scaling relation between $\log(R_{\rm e})$ and magnitude. The merging BCDs (MBCD) data points are shown in  green, red and black colors representing inner, outer and overall sizes respectively.  For the reference, we have used a sample of early-type dwarf galaxies (shown in gray dots) from \citet{Janz08}. The yellow dots represent Virgo cluster BCD studied by \citet{Meyer14}, and we show compact elliptical in a gray square sample taken from \citet{Chilingarian09}}
\label{magsize}
\end{figure}

\begin{figure}
\includegraphics[width=8cm]{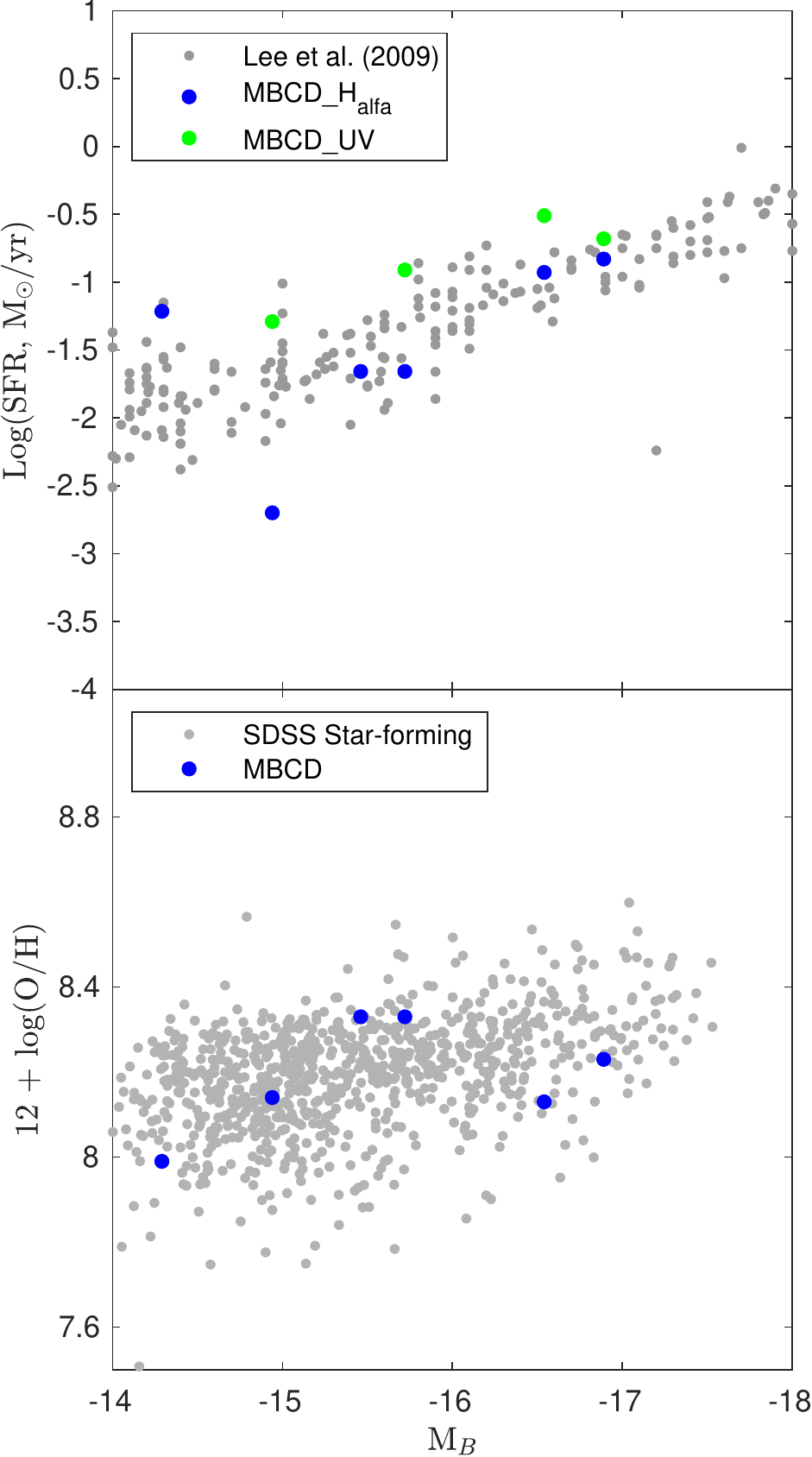}
\caption{\textbf{Top}: Relation between B-band absolute magnitude and $\log$(SFR). The merging BCDs (MBCD) are shown in green and blue symbols, representing SFR derived from UV and H$_{\alpha}$ emission line flux, respectively. We can see only four galaxies for UV because there is no UV data for galaxies D036 and D076. The comparison sample, shown in gray dots, are taken from \citet{Lee09}. \textbf{Bottom}: Relation between emission line metallicity and B-band absolute magnitude. MBCDs are shown in blue dots. In this plot, we used a reference sample of star-forming dwarf galaxies that are taken from \citep{Paudel14}.}
\label{sfrplt}
\end{figure}

In Fig.~\ref{magsize}, we show the relation between $\log$(size) and $r$-band absolute magnitude for our galaxies, and for the reference, we used a sample of early-type dwarf galaxies from \citet{Janz08}. In addition, we show a sample of BCDs of Virgo cluster in yellow color dot, they are taken from the study of \citet{Meyer14} and we show compact ellipticals in a gray square sample taken from \citet{Chilingarian09}. Our sample merging BCDs are shown in three different colors: green, red, and black for the inner, outer, and overall components, respectively. As expected, the BCDs fall well below the relation defined by the reference sample dE. They are slightly larger than cE but significantly smaller in size than dEs. We also notice that the inner component sizes are comparable to the overall size of merging BCDs compared to their outer component size implying that the inner component largely dominates the light distribution. On the other hand, the cE population is the most compact class of object in this diagram. 

\subsection{Spectroscopy}

The SFR$_{\text{H}\alpha}$ is lower compared to the cataloged SFR that is derived from Far-Ultra-Violet (FUV) flux. There are many reasons for this discrepancy. The H$_\alpha$ and FUV fluxes trace different time domains of star formation, the former being significantly shorter than the latter. However, they generally follow the relation between $\log$(SFR) and B-band absolute magnitude [Figure~\ref{sfrplt} top panel]. In addition, while measuring the FUV flux, we considered the entire galaxy using a larger aperture, and the H$\alpha$ fluxes were measured from a central 3$\arcsec$ fiber spectroscopy. The B-band magnitude of our sample galaxies is calculated from the SDSS $g-$band magnitude using the color transformation equation provided by the SDSS webpage\footnote{http://www.sdss3.org/dr8/algorithms/sdssUBVRITransform.php}. The comparison sample in gray dots is taken from \citet{Lee09}, which provides a statistical study of star-formation activity in local volume star-forming galaxies using volume-limited sample.

To derive Oxygen abundances, we employed a method suggested by \citet{Marino13}, where a line ratio between H$_\alpha$ and [NII] is used. Our sample galaxies have a range of $12 + \log$(O/H) between 7.99~dex to 8.33~dex. Being slightly sub-solar, these values are typical of BCDs. In this sample, D076 is the most metal-poor galaxy, and also has the highest gas mass fraction. 

In the bottom panel of Figure~\ref{sfrplt}, we show a relation between B-band absolute magnitude and emission line metallicities. Our BCDs well follow the relation defined by a sample of star-forming dwarf galaxies drawn from the SDSS, which we have taken from \citet{Paudel17}. 
 
To get a rough estimate of the star formation age, we used H$_\alpha$ equivalent width (EW$_{\text{H}\alpha}$). For the majority of our sample, the measured values of EW$_{\text{H}\alpha}$ is larger than 100\AA~, and using the Starburst99 model \citep{Leitherer99} assuming an instantaneous burst of star formation and a half of solar metallicity, these values indicate that the ages of star formation are estimated to be in the order of a few 10~Myr. 

\begin{figure*}
\includegraphics[width=5.9cm]{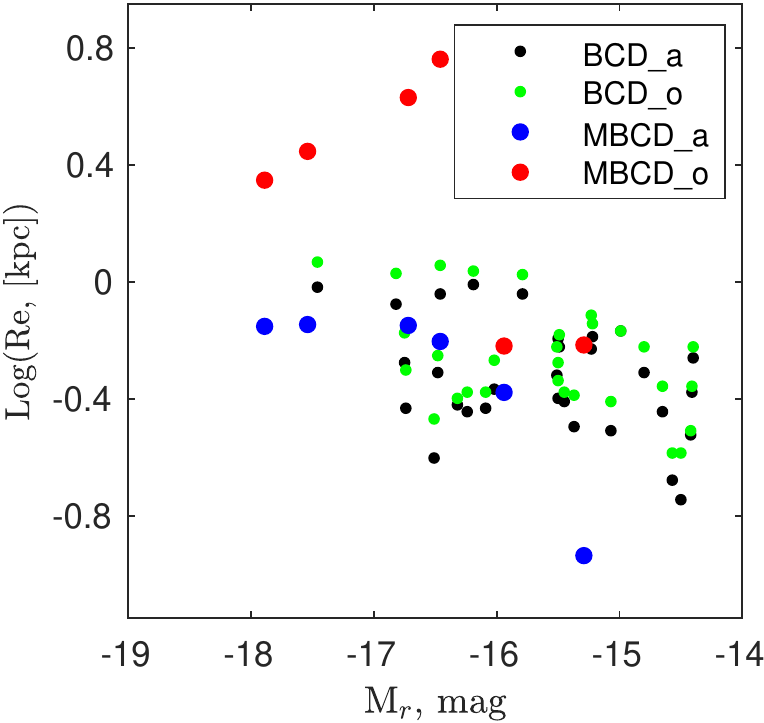}
\includegraphics[width=5.9cm]{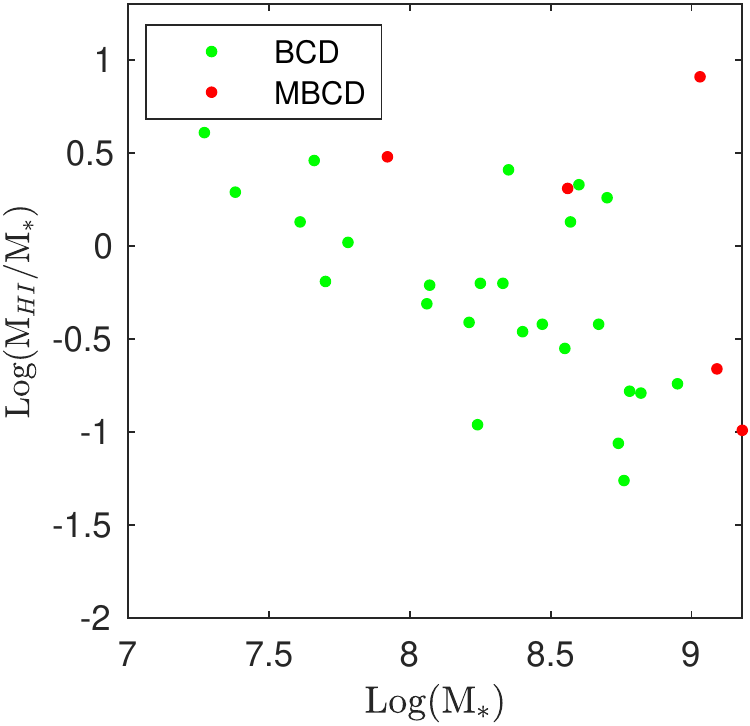}
\includegraphics[width=5.9cm]{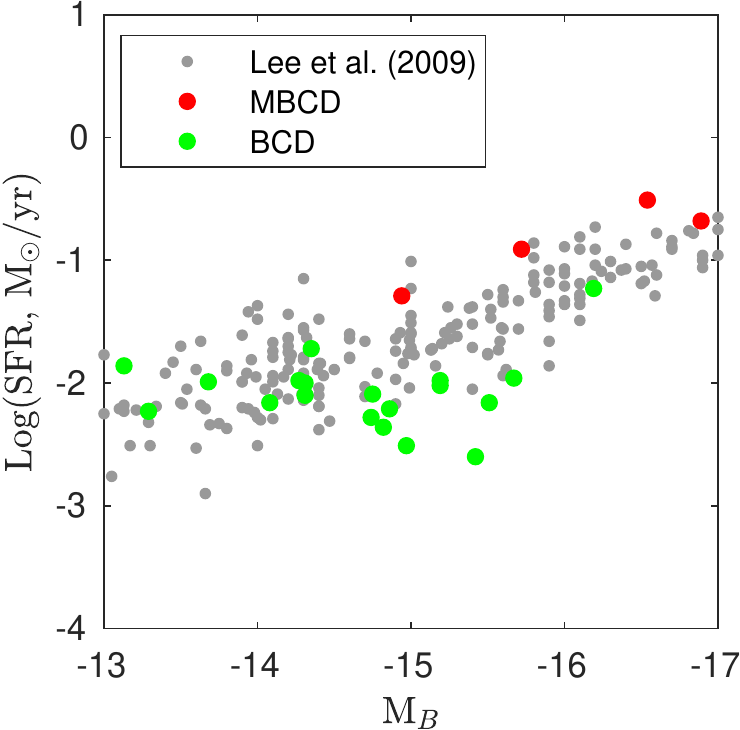}
\caption{Comparison of physical and star-formation parameters between merging BCDs and Virgo cluster BCDs. For the Virgo cluster BCDs, we used \citet{Meyer14} photometric measurement.
 }
\label{mayecmp}
\end{figure*}

\subsection{Gas content}
Queering the CDS catalog service, we find neutral hydrogen (HI) 21-cm radio data for five of our merging BCDs (except D047). They are mainly observed by a single dish telescope (Arecibo observations)  with a typical beam size of 3$\arcmin$ and we expect that the HI flux measurement is for the overall galaxy. Using equation $M_{\text{HI}} = 2.356\times 10^{5}\,d^{2}\int S_{c}(v)~dv\,M_{\sun}$\footnote{where $d$ is the distance in Mpc and $\int S_{c}(v)$ is expressed in Jy and $v$ in km/s \citep{Roberts78}},  we derived HI mass from the 21-cm radio emission flux and we find that our merging BCDs possess an $M_{\text{HI}}$ in the range $2.5\times10^8$~M$_{\sun}$ to $8.7\times10^9$~M$_{\sun}$. The calculated $M_{\text{HI}}$ to blue band luminosity ratio varies between 0.5~M$_{\sun}$/L$_{\sun}$ to 0.8~M$_{\sun}$/L$_{\sun}$ for our sample galaxies. This number is similar to an average value of $M_{\text{HI}}$/$L_{\text{B}}$ for a typical sample of BCDs studied in \citet{Huchtmeier05} but significantly lower than an average value of low-surface brightness dwarf galaxies found in the isolated environments \citep{Pustilnik11}. Note, however, the estimated stellar mass using a single optical color has uncertainty as large as 0.3~dex \citep{Zhang17}. To make a more accurate estimation of stellar mass, a detailed analysis of spectral energy distribution with a longer baseline wavelength is required. 

\section{Discussion and Conclusions}
This work studies a sample of compact merging dwarf galaxies and performs a detailed analysis of morphological parameters. These compact star-forming dwarf galaxies are selected from a catalog of merging dwarf galaxies prepared by \citet{Paudel18a}. Our detailed image analysis shows that, although they are compact systems, these dwarf galaxies' light profiles are no better modeled with one component S\'ersic function. They are better represented with two components: (i) inner compact and (ii) outer extended low-surface brightness component. However, the overall sizes of these galaxies are also significantly smaller compared to the normal dwarf galaxies and the size of the inner component is comparable to the compact early-type galaxies. On the other hand, the relation between B-band magnitude and emission line metallicity and SFR show no anomaly, being consistent with the relation defined by normal star-forming dwarf galaxies. However, a typical 0.3 dex scatter is seen in the relation between SFR and B-band magnitude, even for normal galaxy samples.

\subsection{Comparison to the previous study} 

It is well known that BCDs have two components --inner compact star-forming core and outer low-surface brightness extended halo \citep{Sung84,Loose86}. A detailed morphological property of a complete sample of BCDs of the Virgo cluster was done by \cite{Meyer14}. They decomposed these two components' light profiles and studied their similarity with various morphology dwarf galaxies, such as dwarf elliptical or dwarf irregular. The work, however, makes no comment on the origin of star-burst activity and mainly focuses on the future evolution of BCDs in the cluster environment.

In Figure \ref{mayecmp}, we compare the physical properties of normal BCDs of the Virgo cluster with that of merging BCDs of this study. For the Virgo cluster, we used  \cite{Meyer14} photometric measurement. In the leftmost panel, we show the size magnitude relation of BCDs, and particularly we noticed that although the overall sizes of both sample BCDs are similar, the outer component sizes of merging BCDs are significantly larger than that of normal Virgo BCDs. Such extended outer component, indeed, mainly shows the presence of tidal components that may have built up by merging/accreting other galaxies.

The middle panel of Figure \ref{mayecmp} displays a scaling relation between gas mass fraction and stellar mass. The Virgo BCDs gas masses are estimated from 21 cm neutral Hydrogen emission line flux provided by ALPHA-ALPHA survey \citep{Haynes18} and stellar masses are derived similarly to that derived for merging BCDs from $r-$band magnitude and $g-r$ color. It reveals that merging BCDs tends to have a higher gas mass fraction than BCDs of the Virgo cluster. The discrepancy in gas mass fraction can also be attributed to their environmental differences. Our merging BCDs are primarily located in field environments where galaxies are dominantly gas-rich and star-forming. In the rightmost panel, we show a relation between B-band absolute magnitude and Log(SFR), and similar to Figure \ref{sfrplt}, we added \cite{Lee09} data for the reference. The star-formation rates of Virgo BCDs are estimated from FUV flux provided by the GALEX UV survey of Virgo cluster \citep{Boselli11}. As we have already noticed in Figure \ref{sfrplt}, merging BCDs clearly shows enhanced star-formation activity compared to the reference star-formation sequence provided by  \cite{Lee09}, and Virgo BCDs are located below the sequence.

\subsection{Formation of BCDs through the merger} 
Various observational and theoretical lines of evidence suggest that the evolutionary pathways of BCDs have to be diverse, and multitudes of starburst mechanisms are discussed. Particularly, dwarf-dwarf mergers, interactions, or gas inflow may trigger the gas collapse or strong dissipation, which ignites the burst of star formation in these low-mass galaxies \citep{Papaderos96,Nusser05,Bournaud09,Elmegreen12,Zhang20a,Nusser22}.

\citet{Zhang20a,Zhang20b} studied a link between a dwarf-dwarf merger event and enhanced star formation in a Virgo cluster BCD, VCC\,848. Using a combination of JVLA HI emission line map and optical imaging, they examined the effect of the merger on the stellar and the gaseous distributions and concluded that although VCC\,848 possesses a central burst of star-formation, it has less than 30\% of gas concentration in the centre.  Comparing the general morphological feature with $N$-body hydrodynamical simulations results, they suggested the timing of the merger about a Giga year ago. On contrary to VCC\,848 environments, our merging BCDs are located in a low, dense environment.

While studying the HI morphology of a sample of starburst dwarfs (selected from BCD-like galaxies), \cite{Lelli14} found that the outer extended HI part was more asymmetric in star-burst dwarfs than normal star-forming galaxies, and the age of star-burst is correlated with the strength of asymmetric feature, suggesting that some external mechanism triggered the starburst. In a similar line, \cite{Ramya11} suggests that some of the BCDs they studied might have a tidal origin.

Several physical properties of our sample merging dwarf galaxies, i.e., color, metal content and star-formation rate are fairly similar to a typical BCD and there is little doubt that the star formation activity is affected by the interaction. The star-formation ages derived from H$_\alpha$ equivalent width reveal that the burst of star-formation is a recent activity and the event is very likely triggered by the interaction. Galaxy merger efficiently supplies the gas towards the center helping to ignite a central starburst \citep{Bekki08} and such merger-induced central starburst builds a central blue core that dominates the optical light.

How would these compact BCDs evolve after ceasing the star-formation activity? The sizes of star-forming galaxies are likely to increase as the dynamic and stellar population evolution proceeds, which leads to a decrease in the surface brightness \citep{Wellons15}. These galaxies are significantly compact compared to the normal galaxy. The current star-formation rate is not significantly high and the SDSS spectrum reveals strong higher-order Balmer absorption feature, indicating that the star-formation activity is a temporal feature and it hosts a significant amount of old stellar population. The presence of old stellar population could be due to the effect of the accretion of a smaller satellite by these galaxies and the event eventually perturbed the gaseous disk of the primary, which leads to a burst of star formation at the center. 

A general prediction of early simulations is that galaxies mergers develop substantial gas inflows and induce intense starbursts at the center producing core-like structure \citep{Mihos94}.  The intense starburst at the center of our merging compact dwarf galaxies appears to align with these simulations' results. It has also been proposed that a pre-existing compact core, i.e., a dense central bulge, may help to maintain an intense starburst activity \citep{Mihos96}.

\section*{Acknowledgements}
D.~N.~Chhatkuli acknowledges the University Grants Commission of Nepal for financial support (Ph. D. -75/76-S \& T-13). S.P. acknowledges support from the New Researcher Program (Shinjin grant No. 2019R1C1C1009600) through the National Research Foundation of Korea. J.Y. was supported by a KIAS Individual Grant (QP089901) via the Quantum Universe Center at Korea Institute for Advanced Study. This study is based on the archival images and spectra from the Sloan Digital Sky Survey (the full acknowledgment can be found at \url{https://www.sdss.org/collaboration/acknowledgements}).

\section{DATA AVAILABILITY}
Most of the data underlying this article are publicly available.
The derived data generated in this research will also be shared on reasonable request to the corresponding author.




\begin{thebibliography}{}
\makeatletter
\relax
\def\mn@urlcharsother{\let\do\@makeother \do\$\do\&\do\#\do\^\do\_\do\%\do\~}
\def\mn@doi{\begingroup\mn@urlcharsother \@ifnextchar [ {\mn@doi@}
  {\mn@doi@[]}}
\def\mn@doi@[#1]#2{\def\@tempa{#1}\ifx\@tempa\@empty \href
  {http://dx.doi.org/#2} {doi:#2}\else \href {http://dx.doi.org/#2} {#1}\fi
  \endgroup}
\def\mn@eprint#1#2{\mn@eprint@#1:#2::\@nil}
\def\mn@eprint@arXiv#1{\href {http://arxiv.org/abs/#1} {{\tt arXiv:#1}}}
\def\mn@eprint@dblp#1{\href {http://dblp.uni-trier.de/rec/bibtex/#1.xml}
  {dblp:#1}}
\def\mn@eprint@#1:#2:#3:#4\@nil{\def\@tempa {#1}\def\@tempb {#2}\def\@tempc
  {#3}\ifx \@tempc \@empty \let \@tempc \@tempb \let \@tempb \@tempa \fi \ifx
  \@tempb \@empty \def\@tempb {arXiv}\fi \@ifundefined
  {mn@eprint@\@tempb}{\@tempb:\@tempc}{\expandafter \expandafter \csname
  mn@eprint@\@tempb\endcsname \expandafter{\@tempc}}}

\bibitem[\protect\citeauthoryear{{Ahn} et~al.,}{{Ahn} et~al.}{2012}]{Ahn12}
{Ahn} C.~P.,  et~al., 2012, \mn@doi [\apjs] {10.1088/0067-0049/203/2/21}, \href
  {http://adsabs.harvard.edu/abs/2012ApJS..203...21A} {203, 21}

\bibitem[\protect\citeauthoryear{{Annibali} et~al.,}{{Annibali}
  et~al.}{2016}]{annibali16}
{Annibali} F.,  et~al., 2016, \mn@doi [\apjl] {10.3847/2041-8205/826/2/L27},
  \href {http://adsabs.harvard.edu/abs/2016ApJ...826L..27A} {826, L27}

\bibitem[\protect\citeauthoryear{{Bekki}}{{Bekki}}{2008}]{Bekki08}
{Bekki} K.,  2008, \mn@doi [\mnras] {10.1111/j.1745-3933.2008.00489.x}, \href
  {http://adsabs.harvard.edu/abs/2008MNRAS.388L..10B} {388, L10}

\bibitem[\protect\citeauthoryear{{Boselli} \& {Gavazzi}}{{Boselli} \&
  {Gavazzi}}{2006}]{Boselli06}
{Boselli} A.,  {Gavazzi} G.,  2006, \mn@doi [\pasp] {10.1086/500691}, \href
  {http://adsabs.harvard.edu/abs/2006PASP..118..517B} {118, 517}

\bibitem[\protect\citeauthoryear{{Boselli} et~al.,}{{Boselli}
  et~al.}{2011}]{Boselli11}
{Boselli} A.,  et~al., 2011, \mn@doi [\aap] {10.1051/0004-6361/201016389},
  \href {https://ui.adsabs.harvard.edu/abs/2011A&A...528A.107B} {528, A107}

\bibitem[\protect\citeauthoryear{{Bournaud} \& {Elmegreen}}{{Bournaud} \&
  {Elmegreen}}{2009}]{Bournaud09}
{Bournaud} F.,  {Elmegreen} B.~G.,  2009, \mn@doi [\apjl]
  {10.1088/0004-637X/694/2/L158}, \href
  {https://ui.adsabs.harvard.edu/abs/2009ApJ...694L.158B} {694, L158}

\bibitem[\protect\citeauthoryear{{Cardelli}, {Clayton}  \& {Mathis}}{{Cardelli}
  et~al.}{1989}]{Cardelli89}
{Cardelli} J.~A.,  {Clayton} G.~C.,   {Mathis} J.~S.,  1989, \mn@doi [\apj]
  {10.1086/167900}, \href
  {https://ui.adsabs.harvard.edu/abs/1989ApJ...345..245C} {345, 245}

\bibitem[\protect\citeauthoryear{{Chen}, {C{\^o}t{\'e}}, {West}, {Peng}  \&
  {Ferrarese}}{{Chen} et~al.}{2010}]{Chen10}
{Chen} C.-W.,  {C{\^o}t{\'e}} P.,  {West} A.~A.,  {Peng} E.~W.,   {Ferrarese}
  L.,  2010, \mn@doi [\apjs] {10.1088/0067-0049/191/1/1}, \href
  {http://adsabs.harvard.edu/abs/2010ApJS..191....1C} {191, 1}

\bibitem[\protect\citeauthoryear{{Chilingarian}, {Cayatte}, {Revaz}, {Dodonov},
  {Durand}, {Durret}, {Micol}  \& {Slezak}}{{Chilingarian}
  et~al.}{2009}]{Chilingarian09}
{Chilingarian} I.,  {Cayatte} V.,  {Revaz} Y.,  {Dodonov} S.,  {Durand} D.,
  {Durret} F.,  {Micol} A.,   {Slezak} E.,  2009, \mn@doi [Science]
  {10.1126/science.1175930}, \href
  {https://ui.adsabs.harvard.edu/abs/2009Sci...326.1379C} {326, 1379}

\bibitem[\protect\citeauthoryear{{Dey} et~al.,}{{Dey} et~al.}{2019}]{Dey19}
{Dey} A.,  et~al., 2019, \mn@doi [\aj] {10.3847/1538-3881/ab089d}, \href
  {https://ui.adsabs.harvard.edu/abs/2019AJ....157..168D} {157, 168}

\bibitem[\protect\citeauthoryear{{Drinkwater} \& {Hardy}}{{Drinkwater} \&
  {Hardy}}{1991}]{Drinkwater91}
{Drinkwater} M.,  {Hardy} E.,  1991, \mn@doi [\aj] {10.1086/115670}, \href
  {https://ui.adsabs.harvard.edu/abs/1991AJ....101...94D} {101, 94}

\bibitem[\protect\citeauthoryear{{Duc} et~al.,}{{Duc} et~al.}{2011}]{Duc11}
{Duc} P.-A.,  et~al., 2011, \mn@doi [\mnras]
  {10.1111/j.1365-2966.2011.19137.x}, \href
  {http://adsabs.harvard.edu/abs/2011MNRAS.417..863D} {417, 863}

\bibitem[\protect\citeauthoryear{{Elmegreen}, {Zhang}  \& {Hunter}}{{Elmegreen}
  et~al.}{2012}]{Elmegreen12}
{Elmegreen} B.~G.,  {Zhang} H.-X.,   {Hunter} D.~A.,  2012, \mn@doi [\apj]
  {10.1088/0004-637X/747/2/105}, \href
  {https://ui.adsabs.harvard.edu/abs/2012ApJ...747..105E} {747, 105}

\bibitem[\protect\citeauthoryear{{Faber} \& {Lin}}{{Faber} \&
  {Lin}}{1983}]{Faber83}
{Faber} S.~M.,  {Lin} D.~N.~C.,  1983, \mn@doi [\apjl] {10.1086/183970}, \href
  {http://adsabs.harvard.edu/abs/1983ApJ...266L..17F} {266, L17}

\bibitem[\protect\citeauthoryear{{Ferrarese} et~al.,}{{Ferrarese}
  et~al.}{2006}]{Ferrarese06}
{Ferrarese} L.,  et~al., 2006, \mn@doi [\apjs] {10.1086/501350}, \href
  {http://adsabs.harvard.edu/abs/2006ApJS..164..334F} {164, 334}

\bibitem[\protect\citeauthoryear{{Geha}, {Guhathakurta}  \& {van der
  Marel}}{{Geha} et~al.}{2005}]{Geha05}
{Geha} M.,  {Guhathakurta} P.,   {van der Marel} R.~P.,  2005, \mn@doi [\aj]
  {10.1086/430188}, \href {http://adsabs.harvard.edu/abs/2005AJ....129.2617G}
  {129, 2617}

\bibitem[\protect\citeauthoryear{{Graham} \& {Guzm{\'a}n}}{{Graham} \&
  {Guzm{\'a}n}}{2003}]{Graham03}
{Graham} A.~W.,  {Guzm{\'a}n} R.,  2003, \mn@doi [\aj] {10.1086/374992}, \href
  {http://adsabs.harvard.edu/abs/2003AJ....125.2936G} {125, 2936}

\bibitem[\protect\citeauthoryear{{Graham}, {Driver}, {Petrosian}, {Conselice},
  {Bershady}, {Crawford}  \& {Goto}}{{Graham} et~al.}{2005}]{Graham05}
{Graham} A.~W.,  {Driver} S.~P.,  {Petrosian} V.,  {Conselice} C.~J.,
  {Bershady} M.~A.,  {Crawford} S.~M.,   {Goto} T.,  2005, \mn@doi [\aj]
  {10.1086/444475}, \href
  {https://ui.adsabs.harvard.edu/abs/2005AJ....130.1535G} {130, 1535}

\bibitem[\protect\citeauthoryear{{Haynes} et~al.,}{{Haynes}
  et~al.}{2018}]{Haynes18}
{Haynes} M.~P.,  et~al., 2018, \mn@doi [\apj] {10.3847/1538-4357/aac956}, \href
  {https://ui.adsabs.harvard.edu/abs/2018ApJ...861...49H} {861, 49}

\bibitem[\protect\citeauthoryear{{Huchtmeier}, {Gopal-Krishna}  \&
  {Petrosian}}{{Huchtmeier} et~al.}{2005}]{Huchtmeier05}
{Huchtmeier} W.~K.,  {Gopal-Krishna}  {Petrosian} A.,  2005, \mn@doi [\aap]
  {10.1051/0004-6361:20041401}, \href
  {http://adsabs.harvard.edu/abs/2005A%26A...434..887H} {434, 887}

\bibitem[\protect\citeauthoryear{{Huxor}, {Phillipps}  \& {Price}}{{Huxor}
  et~al.}{2013}]{Huxor13}
{Huxor} A.~P.,  {Phillipps} S.,   {Price} J.,  2013, \mn@doi [\mnras]
  {10.1093/mnras/stt014}, \href
  {https://ui.adsabs.harvard.edu/abs/2013MNRAS.430.1956H} {430, 1956}

\bibitem[\protect\citeauthoryear{{Janz} \& {Lisker}}{{Janz} \&
  {Lisker}}{2008}]{Janz08}
{Janz} J.,  {Lisker} T.,  2008, \mn@doi [\apjl] {10.1086/595720}, \href
  {http://adsabs.harvard.edu/abs/2008ApJ...689L..25J} {689, L25}

\bibitem[\protect\citeauthoryear{{Jedrzejewski}}{{Jedrzejewski}}{1987}]{Jedrzejewski87}
{Jedrzejewski} R.~I.,  1987, \mn@doi [\mnras] {10.1093/mnras/226.4.747}, \href
  {https://ui.adsabs.harvard.edu/abs/1987MNRAS.226..747J} {226, 747}

\bibitem[\protect\citeauthoryear{{Kennicutt}}{{Kennicutt}}{1998}]{Kennicutt98}
{Kennicutt} Jr. R.~C.,  1998, \mn@doi [\araa] {10.1146/annurev.astro.36.1.189},
  \href {http://adsabs.harvard.edu/abs/1998ARA%26A..36..189K} {36, 189}

\bibitem[\protect\citeauthoryear{{Khochfar} \& {Silk}}{{Khochfar} \&
  {Silk}}{2006}]{Khochfar06}
{Khochfar} S.,  {Silk} J.,  2006, \mn@doi [\apjl] {10.1086/507768}, \href
  {https://ui.adsabs.harvard.edu/abs/2006ApJ...648L..21K} {648, L21}

\bibitem[\protect\citeauthoryear{{Kormendy}, {Fisher}, {Cornell}  \&
  {Bender}}{{Kormendy} et~al.}{2009}]{Kormendy09}
{Kormendy} J.,  {Fisher} D.~B.,  {Cornell} M.~E.,   {Bender} R.,  2009, \mn@doi
  [\apjs] {10.1088/0067-0049/182/1/216}, \href
  {http://adsabs.harvard.edu/abs/2009ApJS..182..216K} {182, 216}

\bibitem[\protect\citeauthoryear{{Lee} et~al.,}{{Lee} et~al.}{2009}]{Lee09}
{Lee} J.~C.,  et~al., 2009, \mn@doi [\apj] {10.1088/0004-637X/706/1/599}, \href
  {http://adsabs.harvard.edu/abs/2009ApJ...706..599L} {706, 599}

\bibitem[\protect\citeauthoryear{{Leitherer} et~al.,}{{Leitherer}
  et~al.}{1999}]{Leitherer99}
{Leitherer} C.,  et~al., 1999, \mn@doi [\apjs] {10.1086/313233}, \href
  {https://ui.adsabs.harvard.edu/abs/1999ApJS..123....3L} {123, 3}

\bibitem[\protect\citeauthoryear{{Lelli}, {Verheijen}  \& {Fraternali}}{{Lelli}
  et~al.}{2014}]{Lelli14}
{Lelli} F.,  {Verheijen} M.,   {Fraternali} F.,  2014, \mn@doi [\mnras]
  {10.1093/mnras/stu1804}, \href
  {http://adsabs.harvard.edu/abs/2014MNRAS.445.1694L} {445, 1694}

\bibitem[\protect\citeauthoryear{{Lisker}}{{Lisker}}{2009}]{Lisker09}
{Lisker} T.,  2009, \mn@doi [Astronomische Nachrichten]
  {10.1002/asna.200911291}, \href
  {http://adsabs.harvard.edu/abs/2009AN....330.1043L} {330, 1043}

\bibitem[\protect\citeauthoryear{{Loose} \& {Thuan}}{{Loose} \&
  {Thuan}}{1986}]{Loose86}
{Loose} H.~H.,  {Thuan} T.~X.,  1986, in Star-forming Dwarf Galaxies and
  Related Objects. pp 73--88

\bibitem[\protect\citeauthoryear{{Marino} et~al.,}{{Marino}
  et~al.}{2013}]{Marino13}
{Marino} R.~A.,  et~al., 2013, \mn@doi [\aap] {10.1051/0004-6361/201321956},
  \href {http://adsabs.harvard.edu/abs/2013A%26A...559A.114M} {559, A114}

\bibitem[\protect\citeauthoryear{{Meyer}, {Lisker}, {Janz}  \&
  {Papaderos}}{{Meyer} et~al.}{2014}]{Meyer14}
{Meyer} H.~T.,  {Lisker} T.,  {Janz} J.,   {Papaderos} P.,  2014, \mn@doi
  [\aap] {10.1051/0004-6361/201220700}, \href
  {https://ui.adsabs.harvard.edu/abs/2014A&A...562A..49M} {562, A49}

\bibitem[\protect\citeauthoryear{{Mihos} \& {Hernquist}}{{Mihos} \&
  {Hernquist}}{1994}]{Mihos94}
{Mihos} J.~C.,  {Hernquist} L.,  1994, \mn@doi [\apjl] {10.1086/187679}, \href
  {https://ui.adsabs.harvard.edu/abs/1994ApJ...437L..47M} {437, L47}

\bibitem[\protect\citeauthoryear{{Mihos} \& {Hernquist}}{{Mihos} \&
  {Hernquist}}{1996}]{Mihos96}
{Mihos} J.~C.,  {Hernquist} L.,  1996, \mn@doi [\apj] {10.1086/177353}, \href
  {https://ui.adsabs.harvard.edu/abs/1996ApJ...464..641M} {464, 641}

\bibitem[\protect\citeauthoryear{{Noeske}, {Iglesias-P{\'a}ramo},
  {V{\'{\i}}lchez}, {Papaderos}  \& {Fricke}}{{Noeske} et~al.}{2001}]{Noeske01}
{Noeske} K.~G.,  {Iglesias-P{\'a}ramo} J.,  {V{\'{\i}}lchez} J.~M.,
  {Papaderos} P.,   {Fricke} K.~J.,  2001, \mn@doi [\aap]
  {10.1051/0004-6361:20010446}, \href
  {http://adsabs.harvard.edu/abs/2001A%26A...371..806N} {371, 806}

\bibitem[\protect\citeauthoryear{{Nusser}}{{Nusser}}{2005}]{Nusser05}
{Nusser} A.,  2005, \mn@doi [\mnras] {10.1111/j.1365-2966.2005.09233.x}, \href
  {https://ui.adsabs.harvard.edu/abs/2005MNRAS.361..977N} {361, 977}

\bibitem[\protect\citeauthoryear{{Nusser} \& {Silk}}{{Nusser} \&
  {Silk}}{2022}]{Nusser22}
{Nusser} A.,  {Silk} J.,  2022, \mn@doi [\mnras] {10.1093/mnras/stab3121},
  \href {https://ui.adsabs.harvard.edu/abs/2022MNRAS.509.2979N} {509, 2979}

\bibitem[\protect\citeauthoryear{{Oser}, {Ostriker}, {Naab}, {Johansson}  \&
  {Burkert}}{{Oser} et~al.}{2010}]{Oser10}
{Oser} L.,  {Ostriker} J.~P.,  {Naab} T.,  {Johansson} P.~H.,   {Burkert} A.,
  2010, \mn@doi [\apj] {10.1088/0004-637X/725/2/2312}, \href
  {https://ui.adsabs.harvard.edu/abs/2010ApJ...725.2312O} {725, 2312}

\bibitem[\protect\citeauthoryear{{Papaderos}, {Loose}, {Thuan}  \&
  {Fricke}}{{Papaderos} et~al.}{1996}]{Papaderos96}
{Papaderos} P.,  {Loose} H.-H.,  {Thuan} T.~X.,   {Fricke} K.~J.,  1996, \aaps,
  \href {http://adsabs.harvard.edu/abs/1996A%26AS..120..207P} {120, 207}

\bibitem[\protect\citeauthoryear{{Paudel} \& {Ree}}{{Paudel} \&
  {Ree}}{2014}]{Paudel14b}
{Paudel} S.,  {Ree} C.~H.,  2014, preprint, \href
  {http://adsabs.harvard.edu/abs/2014arXiv1410.7848P} {} (\mn@eprint {arXiv}
  {1410.7848})

\bibitem[\protect\citeauthoryear{{Paudel} \& {Sengupta}}{{Paudel} \&
  {Sengupta}}{2017}]{Paudel17b}
{Paudel} S.,  {Sengupta} C.,  2017, \mn@doi [\apjl] {10.3847/2041-8213/aa95bf},
  \href {http://adsabs.harvard.edu/abs/2017ApJ...849L..28P} {849, L28}

\bibitem[\protect\citeauthoryear{{Paudel}, {Lisker}, {Hansson}  \&
  {Huxor}}{{Paudel} et~al.}{2014}]{Paudel14}
{Paudel} S.,  {Lisker} T.,  {Hansson} K.~S.~A.,   {Huxor} A.~P.,  2014, \mn@doi
  [\mnras] {10.1093/mnras/stu1171}, \href
  {http://adsabs.harvard.edu/abs/2014MNRAS.443..446P} {443, 446}

\bibitem[\protect\citeauthoryear{{Paudel} et~al.,}{{Paudel}
  et~al.}{2017}]{Paudel17}
{Paudel} S.,  et~al., 2017, \mn@doi [\apj] {10.3847/1538-4357/834/1/66}, \href
  {http://adsabs.harvard.edu/abs/2017ApJ...834...66P} {834, 66}

\bibitem[\protect\citeauthoryear{{Paudel}, {Sengupta}  \& {Yoon}}{{Paudel}
  et~al.}{2018a}]{Paudel18b}
{Paudel} S.,  {Sengupta} C.,   {Yoon} S.-J.,  2018a, \mn@doi [\aj]
  {10.3847/1538-3881/aadb8d}, \href
  {https://ui.adsabs.harvard.edu/abs/2018AJ....156..166P} {156, 166}

\bibitem[\protect\citeauthoryear{{Paudel}, {Smith}, {Yoon},
  {Calder{\'o}n-Castillo}  \& {Duc}}{{Paudel} et~al.}{2018b}]{Paudel18a}
{Paudel} S.,  {Smith} R.,  {Yoon} S.~J.,  {Calder{\'o}n-Castillo} P.,   {Duc}
  P.-A.,  2018b, \mn@doi [\apjs] {10.3847/1538-4365/aad555}, \href
  {https://ui.adsabs.harvard.edu/abs/2018ApJS..237...36P} {237, 36}

\bibitem[\protect\citeauthoryear{{Pearson} et~al.,}{{Pearson}
  et~al.}{2016}]{pearson16}
{Pearson} S.,  et~al., 2016, \mn@doi [\mnras] {10.1093/mnras/stw757}, \href
  {http://adsabs.harvard.edu/abs/2016MNRAS.459.1827P} {459, 1827}

\bibitem[\protect\citeauthoryear{{Penny}, {Pimbblet}, {Conselice}, {Brown},
  {Gr{\"u}tzbauch}  \& {Floyd}}{{Penny} et~al.}{2012}]{Penny12}
{Penny} S.~J.,  {Pimbblet} K.~A.,  {Conselice} C.~J.,  {Brown} M.~J.~I.,
  {Gr{\"u}tzbauch} R.,   {Floyd} D.~J.~E.,  2012, \mn@doi [\apjl]
  {10.1088/2041-8205/758/2/L32}, \href
  {http://adsabs.harvard.edu/abs/2012ApJ...758L..32P} {758, L32}

\bibitem[\protect\citeauthoryear{{Privon} et~al.,}{{Privon}
  et~al.}{2017}]{privon17}
{Privon} G.~C.,  et~al., 2017, \mn@doi [\apj] {10.3847/1538-4357/aa8560}, \href
  {http://adsabs.harvard.edu/abs/2017ApJ...846...74P} {846, 74}

\bibitem[\protect\citeauthoryear{{Pustilnik} \& {Tepliakova}}{{Pustilnik} \&
  {Tepliakova}}{2011}]{Pustilnik11}
{Pustilnik} S.~A.,  {Tepliakova} A.~L.,  2011, \mn@doi [\mnras]
  {10.1111/j.1365-2966.2011.18733.x}, \href
  {http://adsabs.harvard.edu/abs/2011MNRAS.415.1188P} {415, 1188}

\bibitem[\protect\citeauthoryear{{Pustilnik}, {Kniazev}, {Lipovetsky}  \&
  {Ugryumov}}{{Pustilnik} et~al.}{2001}]{Pustilnik01}
{Pustilnik} S.~A.,  {Kniazev} A.~Y.,  {Lipovetsky} V.~A.,   {Ugryumov} A.~V.,
  2001, \mn@doi [\aap] {10.1051/0004-6361:20010555}, \href
  {http://adsabs.harvard.edu/abs/2001A%26A...373...24P} {373, 24}

\bibitem[\protect\citeauthoryear{{Ramya}, {Kantharia}  \& {Prabhu}}{{Ramya}
  et~al.}{2011}]{Ramya11}
{Ramya} S.,  {Kantharia} N.~G.,   {Prabhu} T.~P.,  2011, \mn@doi [\apj]
  {10.1088/0004-637X/728/2/124}, \href
  {https://ui.adsabs.harvard.edu/abs/2011ApJ...728..124R} {728, 124}

\bibitem[\protect\citeauthoryear{{Rich}, {Collins}, {Black}, {Longstaff},
  {Koch}, {Benson}  \& {Reitzel}}{{Rich} et~al.}{2012}]{Rich12}
{Rich} R.~M.,  {Collins} M.~L.~M.,  {Black} C.~M.,  {Longstaff} F.~A.,  {Koch}
  A.,  {Benson} A.,   {Reitzel} D.~B.,  2012, \mn@doi [\nat]
  {10.1038/nature10837}, \href
  {http://adsabs.harvard.edu/abs/2012Natur.482..192R} {482, 192}

\bibitem[\protect\citeauthoryear{{Roberts}}{{Roberts}}{1978}]{Roberts78}
{Roberts} M.~S.,  1978, \mn@doi [\aj] {10.1086/112287}, \href
  {https://ui.adsabs.harvard.edu/abs/1978AJ.....83.1026R} {83, 1026}

\bibitem[\protect\citeauthoryear{{Sarzi} et~al.,}{{Sarzi}
  et~al.}{2006}]{Sarzi06}
{Sarzi} M.,  et~al., 2006, \mn@doi [\mnras] {10.1111/j.1365-2966.2005.09839.x},
  \href {http://adsabs.harvard.edu/abs/2006MNRAS.366.1151S} {366, 1151}

\bibitem[\protect\citeauthoryear{{Sersic}}{{Sersic}}{1968}]{Sersic68}
{Sersic} J.~L.,  1968, {Atlas de Galaxias Australes}

\bibitem[\protect\citeauthoryear{{Stierwalt}, {Besla}, {Patton}, {Johnson},
  {Kallivayalil}, {Putman}, {Privon}  \& {Ross}}{{Stierwalt}
  et~al.}{2015}]{stierwalt15}
{Stierwalt} S.,  {Besla} G.,  {Patton} D.,  {Johnson} K.,  {Kallivayalil} N.,
  {Putman} M.,  {Privon} G.,   {Ross} G.,  2015, \mn@doi [\apj]
  {10.1088/0004-637X/805/1/2}, \href
  {http://adsabs.harvard.edu/abs/2015ApJ...805....2S} {805, 2}

\bibitem[\protect\citeauthoryear{{Straughn} et~al.,}{{Straughn}
  et~al.}{2015}]{Straughn15}
{Straughn} A.~N.,  et~al., 2015, \mn@doi [\apj] {10.1088/0004-637X/814/2/97},
  \href {https://ui.adsabs.harvard.edu/abs/2015ApJ...814...97S} {814, 97}

\bibitem[\protect\citeauthoryear{{Sung}, {Han}, {Ryden}, {Chun}  \&
  {Kim}}{{Sung} et~al.}{1998}]{Sung84}
{Sung} E.-C.,  {Han} C.,  {Ryden} B.~S.,  {Chun} M.-S.,   {Kim} H.-I.,  1998,
  \mn@doi [\apj] {10.1086/305633}, \href
  {https://ui.adsabs.harvard.edu/abs/1998ApJ...499..140S} {499, 140}

\bibitem[\protect\citeauthoryear{{Toloba} et~al.,}{{Toloba}
  et~al.}{2014}]{Toloba14}
{Toloba} E.,  et~al., 2014, \mn@doi [\apj] {10.1088/0004-637X/783/2/120}, \href
  {http://adsabs.harvard.edu/abs/2014ApJ...783..120T} {783, 120}

\bibitem[\protect\citeauthoryear{{Tremonti} et~al.,}{{Tremonti}
  et~al.}{2004}]{Tremonti04}
{Tremonti} C.~A.,  et~al., 2004, \mn@doi [\apj] {10.1086/423264}, \href
  {http://adsabs.harvard.edu/abs/2004ApJ...613..898T} {613, 898}

\bibitem[\protect\citeauthoryear{{Wellons} et~al.,}{{Wellons}
  et~al.}{2015}]{Wellons15}
{Wellons} S.,  et~al., 2015, \mn@doi [\mnras] {10.1093/mnras/stv303}, \href
  {https://ui.adsabs.harvard.edu/abs/2015MNRAS.449..361W} {449, 361}

\bibitem[\protect\citeauthoryear{{Zhang}, {Puzia}  \& {Weisz}}{{Zhang}
  et~al.}{2017}]{Zhang17}
{Zhang} H.-X.,  {Puzia} T.~H.,   {Weisz} D.~R.,  2017, \mn@doi [\apjs]
  {10.3847/1538-4365/aa937b}, \href
  {https://ui.adsabs.harvard.edu/abs/2017ApJS..233...13Z} {233, 13}

\bibitem[\protect\citeauthoryear{{Zhang} et~al.,}{{Zhang}
  et~al.}{2020a}]{Zhang20b}
{Zhang} H.-X.,  et~al., 2020a, \mn@doi [\apjl] {10.3847/2041-8213/ab7825},
  \href {https://ui.adsabs.harvard.edu/abs/2020ApJ...891L..23Z} {891, L23}

\bibitem[\protect\citeauthoryear{{Zhang} et~al.,}{{Zhang}
  et~al.}{2020b}]{Zhang20a}
{Zhang} H.-X.,  et~al., 2020b, \mn@doi [\apj] {10.3847/1538-4357/abab96}, \href
  {https://ui.adsabs.harvard.edu/abs/2020ApJ...900..152Z} {900, 152}

\makeatother
\end{thebibliography}

\bsp	
\label{lastpage}
\end{document}